\documentclass[traditabstract]{aa} 
\usepackage{natbib}
\bibpunct{(}{)}{;}{a}{}{,}

\usepackage{graphicx}
\usepackage{txfonts}
\usepackage{url}

\def\ms{\,m\,s$^{-1}$}         
\def\kms{\,km\,s$^{-1}$}       
\def\msol{$M_\odot$}		
\def\rsol{$R_\odot$}		
\def\denssol{$\rho_\odot$}	
\def\mstar{$M_*$}		
\def\rstar{$R_*$}		
\def\densstar{$\rho_*$}		
\def\mplanet{$M_{\rm pl}$}	
\def\rplanet{$R_{\rm pl}$}	
\def\densplanet{$\rho_{\rm pl}$}
\def\mjup{$M_{\rm Jup}$}	
\def\rjup{$R_{\rm Jup}$}	
\def\densjup{$\rho_{\rm Jup}$}	
\def\teql{$T_{\rm eql}$}
\def\teff{$T_{\rm eff}$}
\def\feh{[Fe/H]}
\def\logg{$\log g_*$}
\def\vsini{$v \sin I$}
\def\mictrb{$\xi_{\rm t}$}
\def\mactrb{$v_{\rm mac}$}

\def\halpha{$H_\alpha$}
\def\ali{$A_{\rm Li}$}		
\def\kms{km\, s$^{-1}$}

\def\secos{$\sqrt{e} \cos \omega$}
\def\sesin{$\sqrt{e} \sin \omega$}

\begin{document}

\title{WASP-31b: a low-density planet transiting a metal-poor, late-F-type dwarf 
star\thanks{
Based in part on observations made with the HARPS spectrograph on the 3.6-m ESO 
telescope (proposal 085.C-0393) and with the CORALIE spectrograph and the Euler  
camera on the 1.2-m Euler Swiss telescope, both at the ESO La Silla Observatory, 
Chile.}
$^{, }$
\thanks{The photometric time-series and radial-velocity data used in this 
work are only available in electronic form at the CDS via anonymous ftp to 
cdsarc.u-strasbg.fr (130.79.128.5) or via 
http://cdsarc.u-strasbg.fr/viz-bin/qcat?J/A+A/531/A60}}

\titlerunning{}

\author{D.~R.~Anderson
        \inst{1}
	\and
A.~Collier~Cameron
        \inst{2}
	\and 
C.~Hellier
        \inst{1}
	\and 
M.~Lendl
        \inst{3}
	\and 
T.~A.~Lister
        \inst{4}
	\and 
P.~F.~L.~Maxted
        \inst{1}
	\and 
D.~Queloz
        \inst{3}
	\and 
B.~Smalley
        \inst{1}
	\and 
A.~M.~S.~Smith
        \inst{1}
	\and 
A.~H.~M.~J.~Triaud
        \inst{3}
	\and 
R.~G.~West
        \inst{5}
	\and
D.~J.~A.~Brown
        \inst{2}
	\and 
M.~Gillon
        \inst{6}
	\and 
F.~Pepe
        \inst{3}
	\and
D.~Pollacco
        \inst{7}
	\and 
D.~S\'egransan
        \inst{3}
	\and
R.~A.~Street
        \inst{4}
	\and
S.~Udry
        \inst{3}
}

\institute{Astrophysics Group, Keele University, Staffordshire ST5 5BG, UK\\
           \email{dra@astro.keele.ac.uk}
           \and
           SUPA, School of Physics and Astronomy, University of St. Andrews, 
           North Haugh, Fife KY16 9SS, UK
	   \and
           Observatoire de Gen\`eve, Universit\'e de Gen\`eve, 51 Chemin 
           des Maillettes, 1290 Sauverny, Switzerland
	   \and
           Las Cumbres Observatory, 6740 Cortona Dr. Suite 102, Santa 
           Barbara, CA 93117, USA
	   \and
           Department of Physics and Astronomy, University of Leicester, 
           Leicester LE1 7RH, UK
           \and
           Institut d'Astrophysique et de G\'eophysique,  Universit\'e de 
           Li\`ege,  All\'ee du 6 Ao\^ut, 17,  Bat.  B5C, Li\`ege 1, Belgium
	   \and
           Astrophysics Research Centre, School of Mathematics \& Physics,
           Queen's University, University Road, Belfast BT7 1NN, UK
}

\date{Received 25 November 2010 / Accepted 18 May 2011}
\authorrunning{D. R. Anderson et al.}
\titlerunning{WASP-31b: a low-density, transiting planet}

 
\abstract
{
We report the discovery of the low-density, transiting giant planet WASP-31b. 
The planet is 0.48 Jupiter masses and 1.55 Jupiter radii. It is in a 3.4-day 
orbit around a metal-poor, late-F-type, $V = 11.7$ dwarf star, which is a 
member of a common proper motion pair. 
In terms of its low density, WASP-31b is second only to WASP-17b, which is a 
more highly irradiated planet of similar mass.
}


\maketitle

%

\section{Introduction}
To date, 107 transiting extrasolar planets have been 
discovered\footnote{2010 Nov 25, http://exoplanet.eu}, the majority of which are 
gas giants in short orbits. 
The radii of a subset of these exoplanets are larger than predicted by standard 
models of irradiated gas giants 
\citep[e.g.,][]{2007ApJ...661..502B,2007ApJ...659.1661F}, including 
TrES-4b \citep{2007ApJ...667L.195M,2009ApJ...691.1145S}, 
WASP-12b \citep{2009ApJ...693.1920H}, and 
WASP-17b \citep{2010ApJ...709..159A,2011MNRAS.416.2108A}.
A number of mechanisms have been proposed as potential solutions to the radius 
anomaly (see \citet{2010exop.book..397F} for a review), each of which involves 
either injecting heat into the planet from an external source or slowing heat 
loss from the planet.

One such mechanism is the dissipation of energy within a planet as heat during 
the tidal circularisation of an eccentric orbit \citep{2001ApJ...548..466B, 
2003ApJ...588..509G, 2008ApJ...681.1631J, 2009ApJ...700.1921I}. 
Such studies suggest that tidal heating may be sufficient to explain the 
large radii of even the most bloated exoplanets, though we would have to be 
observing some systems at very special times. 
A high heating rate, as suggested by \citet{2010A&A...516A..64L}, would mean 
most tidal energy is radiated away by the age typical of the very most bloated 
planets (a few Gyr) and so could not have played a significant role in their 
observed bloating.
However, the current uncertainty in tidal theory allows for a wide range of 
heating rates \citep[e.g.][]{2011ApJ...727...75I}.
Though most studies have considered a transient phase of tidal 
heating, ongoing tidal heating \citep[e.g.][]{2010ApJ...713..751I} would occur 
if an additional companion continues to excite the orbital eccentricity of the 
bloated planet \citep[e.g.][]{2007MNRAS.382.1768M}.

\citet{2007ApJ...661..502B} proposed that enhanced opacities would retard 
the loss of internal heat and thus slow contraction of bloated planets. 
They suggested that enhanced opacities may arise due to the strong optical and 
UV irradiation of short-orbit, gas giants that could alter their atmospheres, 
producing thick hazes, absorbing clouds and non-equilibrium chemical species 
(e.g. tholins or polyacetylenes).

The bloated planets are all very strongly irradiated by their host stars, and a 
small fraction of stellar insolation energy would be sufficient to account for 
the observed degrees of bloating.
\citet{2002A&A...385..156G} suggested that the kinetic energy of strong winds, 
induced in the atmosphere by the large day-night temperature contrasts that 
result from tidal locking, may be transported downward and deposited as thermal 
energy in the deep interior. However, a mechanism to convert the kinetic energy 
into thermal energy would still be required. 
\citet{2010ApJ...725.1146L} and \citet{2010ApJ...721.1113Y} found that 
turbulence is efficient at dissipating kinetic energy. 
Magnetic drag on weakly ionized winds \citep{2010ApJ...719.1421P} and Ohmic 
heating \citep{2010ApJ...714L.238B} are alternative mechanisms.
The non-bloated planets are also highly irradiated. Hence, such a 
mechanism would either have to act more efficiently on the bloated planets, or 
some other property must counteract its effect. 
One such possibility is the presence of a massive core. Indeed, 
\citet{2006A&A...453L..21G} and \citet{2007ApJ...661..502B} found a correlation 
between the core masses required to reproduce the observed radii of known 
exoplanets and the metallicities of their host stars.

In this paper, we present the discovery of the bloated, transiting, giant planet 
WASP-31b. Compared to the ensemble of known short-period planets, 
WASP-31b is moderately irradiated by its low-metallicity host star.


\section{Observations}
\begin{figure}
\includegraphics[width=84mm]{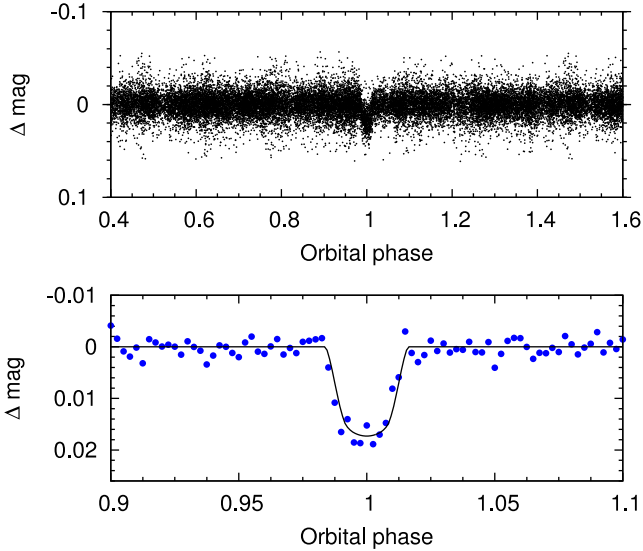} 
\caption{WASP-South discovery light curve. 
\textbf{\emph{Upper panel}}: Photometry folded on the orbital period of 
$P$~=~3.4 d. 
Points with error above three times the median error (0.012 mag) were clipped 
for display purposes. 
\textbf{\emph{Lower panel}}: Photometry folded on the orbital period and binned 
in phase ($\Delta \phi = 0.025$), with the transit model generated from the 
parameters of Table~\ref{tab:mcmc} superimposed.
\label{fig:disc-phot}} 
\end{figure}

WASP-31 is a $V = 11.7$, F7--8 dwarf star located in the constellation Crater. 
WASP-31 has been observed by WASP-South \citep{2006PASP..118.1407P} 
during the first five months of each year since the start of full-scale 
operations (2006 May 4). 
A transit search \citep{2006MNRAS.373..799C} of the resulting 24\,614 usable 
photometric measurements (Figure~\ref{fig:disc-phot}) found a strong, 3.4-d 
periodicity.  

WASP-31 is a visual double with a $V \sim 15.8$ star (2MASS 
11174477-1903521) approximately 35\arcsec\ away. 
The 2MASS colours of the companion suggest that it is a mid-to-late K-type star. 
The proper motions for the two stars listed in the PPMXL 
\citep{2010AJ....139.2440R} and UCAC3 \citep{2010AJ....139.2184Z}
catalogues suggest that this is a common proper motion pair
(Table~\ref{tab:props}). 
The companion is blended with WASP-31 in the WASP images, so we corrected the 
WASP photometry for this contamination to prevent dilution of the transit.

\begin{table}
\caption{Proper motions of WASP-31 and its visual companion
\label{tab:props}}
\begin{tabular}{llll}
\hline
Star		& Catalogue 	& $\mu_{\rm RA}$ (mas) & $\mu_{\rm Dec}$ (mas)	\\
\hline
WASP-31		& UCAC3		& $-$28.2 $\pm$ 1.3 & $-$0.4 $\pm$ 1.6		\\
WASP-31		& PPXML		& $-$25.0 $\pm$ 2.3 & $-$0.1 $\pm$ 2.4 		\\
Companion	& UCAC3		& $-$33.1 $\pm$ 3.1 & +1.0 $\pm$ 3.8		\\
Companion	& PPXML		& $-$28.5 $\pm$ 4.2 & +1.6 $\pm$ 4.2 		\\
\hline
\end{tabular}
\end{table}

Using the CORALIE spectrograph mounted on the 1.2-m Euler-Swiss telescope 
\citep{1996A&AS..119..373B,2000A&A...354...99Q}, we obtained 34 
spectra of WASP-31 during 2009 and a further 13 spectra during 2010. 
As pressure variations can cause CORALIE to drift on short timescales, we 
calibrated the stellar spectra by obtaining simultaneous spectra of a 
thorium-argon lamp. 
In April 2010, we obtained an additional 10 spectra with the HARPS spectrograph 
mounted on the 3.6-m ESO telescope. 
As HARPS is stable at the  1 m s$^{-1}$ night$^{-1}$ level, we obtained a 
calibration at the start of each night of observations. This avoids 
contamination of the stellar spectra by the thorium-argon lamp.
The CORALIE measurement taken at BJD = 2\,454\,971.548671 and the HARPS 
measurement taken at BJD = 2\,455\,299.716991 were both affected by cloud 
cover.

\begin{figure}
\includegraphics[width=84mm]{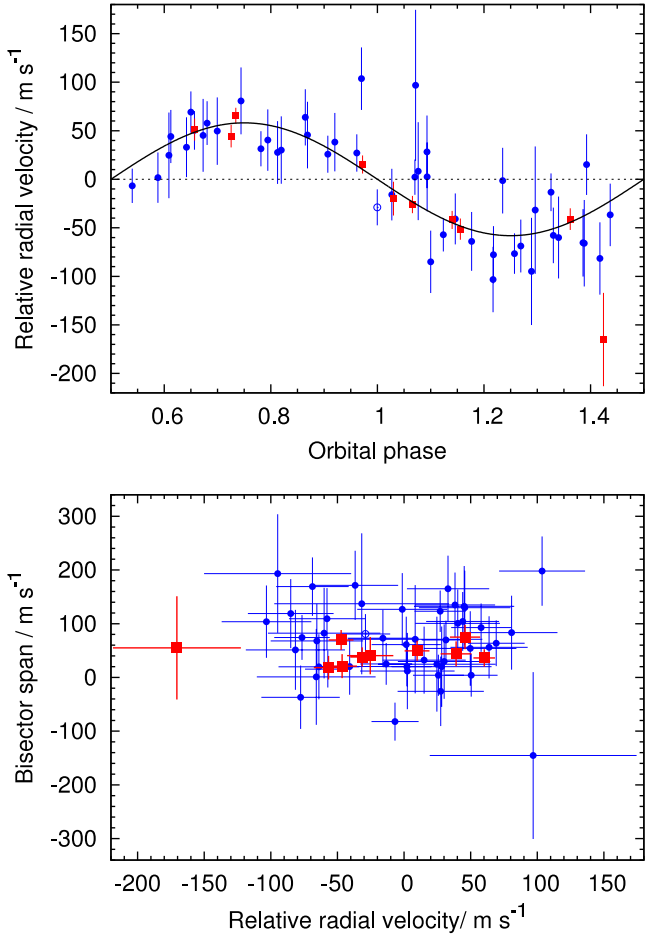}
\caption{
\textbf{\emph{Upper panel}}: Spectroscopic orbit of WASP-31, as illustrated by 
radial velocities from CORALIE (blue circles) and HARPS (red squares). 
The best-fitting Keplerian model, generated from the parameters of 
Table~\ref{tab:mcmc}, is overplotted as a solid line. 
An RV taken at BJD = 2\,455\,168.8468, depicted in the plot by an open circle, 
fell during transit. As we did not treat the Rossiter-McLaughlin 
effect \citep[e.g.,][]{2000A&A...359L..13Q}, we excluded this measurement from 
our combined analysis.
\textbf{\emph{Lower panel}}: A lack of correlation between bisector spans and 
radial velocities rules out a blended eclipsing binary or starspots as the cause 
of the photometric and spectroscopic variaitions. 
We adopted uncertainties on the bisector spans twice the size of those on the 
radial velocities. 
For both plots, the centre-of-mass velocity, $\gamma = -124.92$ \ms, was 
subtracted from the radial velocities and the Keplerian model. 
\label{fig:rv}} 
\end{figure}

The typical signal-to-noise ratio (S/N) per pixel at 550 nm is 18 for the 
CORALIE spectra, with exposure times of 30 min, and 26 for the HARPS spectra, 
with exposure times of 15 min.
Radial-velocity (RV) measurements were computed by weighted cross-correlation 
\citep{1996A&AS..119..373B,2005Msngr.120...22P} with a numerical G2-spectral 
template. 
RV variations were detected with the same period found from the WASP photometry 
and with a semi-amplitude of 58 m s$^{-1}$, consistent with a planetary-mass  
companion. The RV measurements are plotted in the upper panel of 
Figure~\ref{fig:rv}.

To test the hypothesis that the RV variations are due to spectral line 
distortions caused by the presence of cool stellar spots or a blended eclipsing 
binary, a line-bisector analysis \citep{2001A&A...379..279Q,2002A&A...392..215S} 
of the CORALIE and HARPS cross-correlation functions was performed. 
The lack of correlation between bisector span and RV 
(Figure~\ref{fig:rv}, lower panel), especially for the high-precision HARPS 
measurements, supports the identification of the transiting body as a planet.
As an additional test for spots or a blend, we also computed CORALIE RVs by 
weighted cross-correleation with a numerical K5-spectral template. 
The amplitude and phase of the RV variations are the same within errors, 
irrespective of the choice of cross-correlation mask, as would be expected if 
the variations were caused by the presence of a planet 
\citep[e.g.][]{2008A&A...489L...9H}.

To refine the system paramters, we obtained high-S/N transit photometry.
Photometric follow-up observations of WASP-31 were obtained with the 
LCOGT\footnote{http://lcogt.net} 2.0-m Faulkes Telescope North (FTN) on Mt. 
Haleakala, Maui on the night of 2010 Feb 26. The fs03 Spectral Instruments 
camera was used with a $2\times 2$ binning mode, giving a field of view of 
$10\arcmin \times 10\arcmin$ and a pixel scale of 0.303\arcsec\ pixel$^{-1}$. 
The data were taken through a Pan-STARRS $z$ filter, with the telescope 
defocussed to minimise flat-fielding errors and to allow 60-s exposure times 
to be used without saturating.

\begin{figure}
\includegraphics[width=84mm]{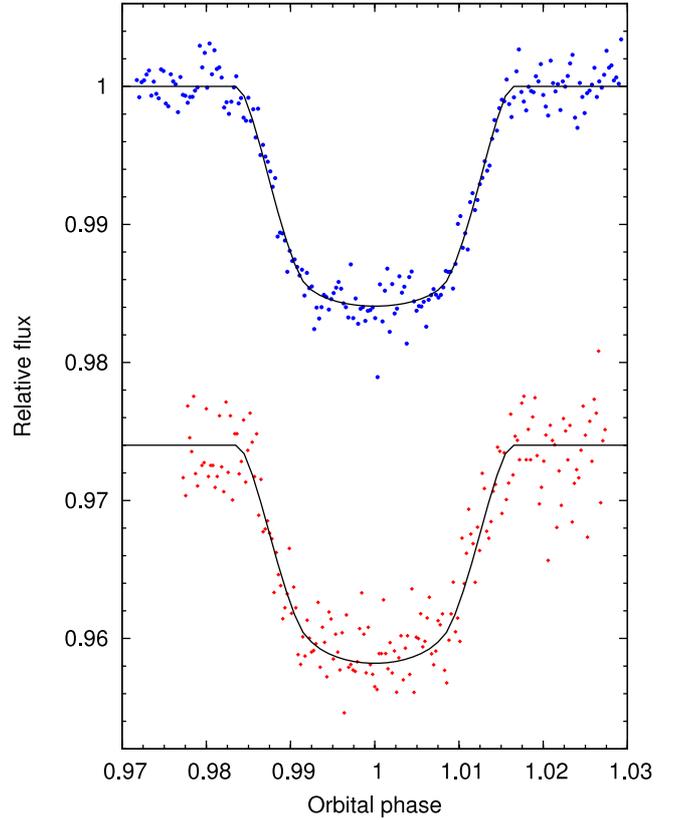} 
\caption{High-S/N transit light curves. The upper observations (blue 
circles) were obtained by FTN, using a Pan-STARRS $z$ filter, on 2010 Feb 26. 
The lower observations (red diamonds), offset in relative flux by 0.026 for 
display, were obtained by Euler, using a Gunn $r$ filter, on 2010 Apr 15. 
The best-fitting transit models generated from the parameters of 
Table~\ref{tab:mcmc} are overplotted. 
\label{fig:fup-phot}}
\end{figure}

The data were pre-processed using the WASP Pipeline \citep{2006PASP..118.1407P} 
to perform masterbias and flat construction, debiassing and flatfielding. Due to 
the very low dark current of the fs03 Fairchild CCD ($<0.0001$ e$^-$ pix$^{-1}$ 
s$^{-1}$), dark subtraction was not performed. 
Aperture photometry was performed using 
DAOPHOT within the IRAF environment using an aperture with a radius of 11 
pixels. Differential photometry was then performed relative to 20 comparison 
stars that were within the FTN field of view (Figure~\ref{fig:fup-phot}).
The RMS of the residuals about the best-fitting model 
(Section~\ref{sys-par}) was 1.4 mmag.

On 2010 April 15 we obtained 4.1 hours of photometry in the Gunn $r$ filter 
with the CCD camera on the Euler-Swiss telescope, covering from 40 min before 
the start of transit until 55 min after it ended. 
The conditions were variable, with seeing of 0.6--1.7\arcsec\ and an airmass 
range of 1.15--1.34.
Euler now employs absolute tracking to keep the stars on the same pixels during 
a whole transit. 
By identifying point sources in each image and matching them with a catalogue, 
the image centre is calculated. Drifts from the nominal position are then 
corrected by adjusting the telescope pointing between exposures.

After bias-subtracting and flat-fielding the images, we performed aperture 
photometry. The flux was extracted for all stars in the 
field and the final light curve (Figure~\ref{fig:fup-phot}) was obtained by 
differential photometry of the target and a reference source obtained by 
combining the 4  brightest reference stars. 
The RMS of the residuals about the best-fitting model 
(Section~\ref{sys-par}) was 2.6 mmag, which was limited by the number of 
available reference stars.

\section{Stellar parameters}

The individual HARPS spectra of WASP-31 were co-added to produce a
single spectrum with an average S/N of around 100:1. The analysis was performed 
using the methods given in \citet{2009A&A...496..259G}. 
The \halpha\ line was used to determine the
effective temperature (\teff), while the Na {\sc i} D and Mg {\sc i} b lines
were used as surface gravity (\logg) diagnostics. The parameters obtained from
the analysis are listed in Table~\ref{tab:stellar}. The elemental abundances
were determined from equivalent width measurements of several clean and
unblended lines. A value for microturbulence (\mictrb) was determined from
Fe~{\sc i} using the method of \cite{1984A&A...134..189M}. The quoted error 
estimates include that given by the uncertainties in \teff, \logg\ and \mictrb, 
as well as the scatter due to measurement and atomic data uncertainties.

The sky-projected stellar rotation velocity (\vsini) was determined by fitting 
the profiles of several unblended Fe~{\sc i} lines. We assumed a value for 
macroturbulence (\mactrb) of 5.2 $\pm$ 0.3 \kms, based on the tabulation by
\cite{2008oasp.book.....G}, and we used an instrumental FWHM of 0.06 $\pm$ 0.01 
\AA, determined from the telluric lines around 6300\AA.
A best-fitting value of \vsini\ = 7.6 $\pm$ 0.4~\kms\ was obtained. However, 
recent work by \cite{2010MNRAS.405.1907B} suggests a lower value for 
macroturbulence of \mactrb\ = 4.2 $\pm$ 0.3 \kms\ which yields a slightly higher 
\vsini\ = 8.1 $\pm$ 0.4~\kms. We therefore adopt the average of these two 
determinations, \vsini\ = 7.9 $\pm$ 0.6~\kms, with the uncertainty being the 
quadrature addition of the individual uncertainties.
If \mactrb\ = 0 \kms, then a value of \vsini\ = 8.7 $\pm$ 0.4~\kms\ is found, 
which is the upper-limit of the sky-projected rotation velocity.

\begin{table}
\centering
\caption{Stellar parameters from the spectroscopic analysis}
\begin{tabular}{lc}
\hline
Parameter & Value \\
\hline
\teff      & 6300 $\pm$ 100 K \\
\logg      & 4.4 $\pm$ 0.1 (cgs)\\
\mictrb    & 1.4 $\pm$ 0.1 \kms \\
\vsini     & 7.9 $\pm$ 0.6 \kms \\
{[Fe/H]}   &$-$0.20 $\pm$ 0.09\\
{[Na/H]}   &$-$0.24 $\pm$ 0.04 \\
{[Mg/H]}   &$-$0.11 $\pm$ 0.06 \\
{[Si/H]}   &$-$0.13 $\pm$ 0.07 \\
{[Ca/H]}   &$-$0.03 $\pm$ 0.08 \\
{[Sc/H]}   &$-$0.07 $\pm$ 0.06 \\
{[Ti/H]}   &$-$0.10 $\pm$ 0.09 \\
{[V/H]}    &$-$0.16 $\pm$ 0.09 \\
{[Cr/H]}   &$-$0.20 $\pm$ 0.08 \\
{[Mn/H]}   &$-$0.45 $\pm$ 0.11 \\
{[Ni/H]}   &$-$0.25 $\pm$ 0.08 \\
log A(Li)[LTE]  &   2.82 $\pm$ 0.08  \\
log A(Li)[NLTE]  &   2.75 $\pm$ 0.08  \\
\mstar     &  1.15 $\pm$ 0.08 \msol \\
\rstar     &  1.12 $\pm$ 0.15 \rsol \\ 
\hline
R.A. (J2000)	& 11$\rm^{h}$17$\rm^{m}$45.35$\rm^{s}$\\
Dec. (J2000)	& $-19^\circ$03$^{'}$17.3$^{''}$\\
$V_{\rm mag}$	& $11.7 \pm 0.2$\\
$J_{\rm mag}$	& $10.91 \pm 0.02$\\
$H_{\rm mag}$	& $10.71 \pm 0.02$\\
$K_{\rm mag}$	& $10.65 \pm 0.03$\\
USNO-B1.0	& 0709-0239208\\
2MASS 		& 11174536-1903171\\
\hline
\end{tabular}
\label{tab:stellar}
\newline {\bf Note:} NLTE Lithium value using correction of \citet{1994A&A...288..860C}.
Mass and Radius estimate using the
\cite{2010A&ARv..18...67T} calibration.
\end{table}

\section{Combined analysis}
\label{sys-par}
The WASP, FTN and Euler photometry were combined with the CORALIE and HARPS 
radial velocities in a simultaneous Markov-chain Monte Carlo (MCMC) analysis 
\citep{2007MNRAS.380.1230C,2008MNRAS.385.1576P}. 
The transit light curve was modeled using the formulation of 
\citet{2002ApJ...580L.171M} with the assumption that  \rplanet$\ll$\rstar. 
Limb-darkening was accounted for using a four-coefficient nonlinear 
limb-darkening model, using fixed coefficients (Table~\ref{tab:ld}) appropriate 
to the passbands and interpolated in effective temperature, surface gravity and 
metallicity from the tabulations of \citet{2000A&A...363.1081C}. 

\begin{table}
\caption{Limb-darkening coefficients} 
\label{tab:ld} 
\begin{tabular}{lcccc}
\hline
Light curve (band)	& $a_1$		& $a_2$		& $a_3$		& $a_4$	\\
\hline
WASP/Euler ($RC$)	& 0.430 	& 0.488 	& $-$0.254	&    0.020 \\
FTN (Pan-STARRS $z$)	& 0.520		& 0.060 	&    0.100	& $-$0.107 \\
\hline
\end{tabular}
\end{table}

The transit light curve is parameterized by the epoch of midtransit 
$T_{\rm 0}$, the orbital period $P$, the planet-to-star area ratio 
(\rplanet/\rstar)$^2$, the approximate duration of the transit from initial to 
final contact $T_{\rm 14}$, and the impact parameter $b = a \cos i/R_{\rm *}$ 
(the distance, in fractional stellar radii, of the transit chord from the 
star's centre). 
The radial-velocity orbit is parameterized by the stellar reflex velocity 
semi-amplitude $K_{\rm *}$, the systemic velocity $\gamma$, 
and \secos\ and \sesin\ 
\citep{2011ApJ...726L..19A}, where $e$ is orbital 
eccentricity and $\omega$ is the argument of periastron. 

The linear scale of the system depends on the orbital separation $a$ which, 
through Kepler's third law, depends on the stellar mass \mstar. 
At each step in the Markov chain, the latest values of stellar density 
\densstar, effective temperature \teff\ and metallicity \feh\ are input in 
to the empirical mass calibration of \citet{2010A&A...516A..33E} to obtain 
\mstar. 
The shapes of the transit light curves \citep{2003ApJ...585.1038S} and the 
radial-velocity curve constrain \densstar, which combines with \mstar\ to 
give \rstar. 
\teff\ and \feh\ are proposal parameters constrained by Gaussian priors with 
mean values and variances derived directly from the stellar spectra 
(Table~\ref{tab:stellar}). 

As the planet-star area ratio is constrained by the measured transit depth, 
\rplanet\ follows from \rstar. The planet mass \mplanet\ is calculated from 
the measured value of $K_1$ and 
\mstar; the planetary density \densplanet\ and surface gravity $\log g_{\rm pl}$ 
then follow. 
We also calculate the blackbody equilibrium temperature \teql\ (assuming zero 
albedo and efficient redistribution of heat from the planet's presumed 
permanent day side to its night side), the transit ingress and egress durations, 
$T_{\rm 12}$ and $T_{\rm 34}$, and the orbital semi-major axis $a$.

At each step in the MCMC procedure, model transit light curves and 
radial-velocity curves are computed from the proposal parameter values, which are 
perturbed from the previous values by a small, random amount. The $\chi^2$ 
statistic is used to judge the goodness of fit of these models to the data and a 
step is accepted if $\chi^2$ is lower than for the previous step. A step 
with higher $\chi^2$ is accepted with a probability $\exp(-\Delta \chi^2$/2). 
In this way, the parameter space around the optimum solution is thoroughly 
explored. 
To give proper weighting to each transit and RV data set, the uncertainties are 
scaled at the start of the MCMC so as to obtain a reduced $\chi^2$ of unity. 
We allow for a systematic instrumental offset, $\Delta \gamma_{\rm HARPS}$, 
between the CORALIE and HARPS spectrographs.

From an initial MCMC fit for an eccentric orbit, we found 
$e = 0.027^{+0.034}_{-0.020}$, with a 3-$\sigma$ upper limit of 0.13. 
The $F$-test approach of \citet{1971AJ.....76..544L} indicates that there is a 
66\% probability that an eccentricity of or above the fitted value could have 
arisen by chance if the the underlying orbit is in fact circular. 
As such, we impose a circular orbit, but we note that doing so 
has no signicant effect as the fitted eccentricity was so small.

The median values and 1 $\sigma$ uncertainties of the system parameters derived 
from the MCMC model fit are presented in Table~\ref{tab:mcmc}. 
The corresponding best-fitting transit light curves are shown in 
Figure~\ref{fig:disc-phot} and Figure~\ref{fig:fup-phot}, and the best-fitting 
RV curve is shown in Figure~\ref{fig:rv}. 

Though the visual companion is 40 times fainter than WASP-31 and is resolved 
in the Euler and FTN images, we did correct the WASP photometry for the 
contamination prior to producing the MCMC solution presented. We checked the 
effect of the contamination by producing another MCMC solution using the 
non-corrected WASP photometry. The best-fitting parameter values were the same 
to within a tenth of an error bar.

Without exquisite photometry, our implentation of MCMC tends to bias 
the impact parameter, and thus \rstar\ and \rplanet, to 
higher values.  
This is because, with low-S/N photometry, the transit ingress and 
egress durations are uncertain, and symmetric uncertainties in those translate 
into asymmetric uncertainties in $b$ and thus in \rstar. 
The effect on the stellar and planetary radii is larger for 
high-impact-parameter planets such as WASP-31b.
Therefore we explored an MCMC with a main-sequence (MS) prior imposed 
\citep{2007MNRAS.380.1230C}. This employs a Bayesian penalty to ensure that, in 
accepted steps, the values of stellar radius are consistent with the values of 
stellar mass for a main-sequence star. 
The differences between the solutions with and without MS priors are small and 
within errors, indicating that the transit light curves are of 
a quality such that the ingress and egress durations are measured sufficiently 
well. 
As such, we adopt the solution without the MS prior, which has slightly more 
conservative error bars. 

\begin{table} 
\caption{System parameters from the combined analysis} 
\label{tab:mcmc} 
\begin{tabular*}{0.4\textwidth}{@{\extracolsep{\fill}}lc} 
\hline 
Parameter (Unit) & Value\\ 
\hline
$P$ (d) & $3.4059096 \pm 0.000005$\\
$T_{\rm 0}$ (HJD) & $2\,455\,192.6887 \pm 0.0003$\\
$T_{\rm 14}$ (d) & $0.1103 \pm 0.0013$\\
$T_{\rm 12}=T_{\rm 34}$ (d) & $0.0285 \pm 0.0018$\\
$a$/\rstar\ & $8.00 \pm 0.19$\\
$R_{\rm pl}^{2}$/R$_{*}^{2}$ & $0.01615 \pm 0.00027$\\
$b$ & $0.780 \pm 0.013$\\
$i$ ($^\circ$) \medskip & $84.41 \pm 0.22$\\
$K_{\rm 1}$ (\ms) & $58.1 \pm 3.4$\\
$a$ (AU)  & $0.04659 \pm 0.00035$\\
$e$ & 0 (adopted)\\
$\gamma_{\rm CORALIE}$ (\ms) & $-124.923 \pm 0.035$\\
$\Delta \gamma_{\rm HARPS}$ (\ms) \medskip & $-5.268 \pm  0.093$\\
\mstar\ (\msol) & $1.163 \pm 0.026$\\
\rstar\ (\rsol) & $1.252 \pm 0.033$\\
\densstar\ (\denssol) & $0.592 \pm 0.042$\\
\logg\ (cgs) & $4.308 \pm 0.020$\\
\teff\ (K) & $6302 \pm 102$\\
\feh\ \medskip & $-0.200 \pm 0.090$\\
\mplanet\ (\mjup) & $0.478 \pm 0.029$\\
\rplanet\ (\rjup) & $1.549 \pm 0.050$\\
\densplanet\ (\densjup) & $0.129 \pm 0.014$\\
$\log g_{\rm pl}$ (cgs) & $2.659 \pm 0.036$\\
\teql\ (K) & $1575 \pm 32$\\
\hline 
\end{tabular*} 
\end{table}

\section{System age}

\begin{figure}
\includegraphics[width=84mm]{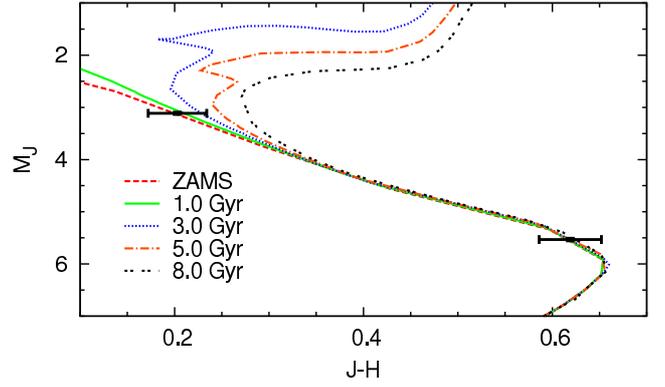}
\caption{Colour-magnitude diagram for WASP-31 and its companion.
Various isochrones from \citet{2008A&A...482..883M} are given, with ages
indicated in the figure. No de-reddening has been applied as $A_{\rm V}=0$.}
\label{fig:cmd}
\end{figure}

Using 2MASS photometry we constructed a colour-magnitude diagram
(Figure~\ref{fig:cmd}) for WASP-31 and its common proper motion, K-type 
companion. 
A distance modulus of $7.8\pm0.2$ ($360\pm30$~pc) 
is required to place the companion on the main-sequence, which puts WASP-31 
between the zero-age main sequence and 1-Gyr age lines. Thus, if 
the two stars are of common origin, they are most probably around 1-Gyr old, 
with an approximate upper age limit of 4 Gyr.
Combining the absolute $V_{\rm mag}$ of an F7--8 star 
\citep{2008oasp.book.....G} with the measured $J-K$ and $H-K$ colours of WASP-31 
from 2MASS and the intrinsic colours from \citet{1983A&A...128...84K}, we 
derive an interstellar extinction $A_{\rm V} = 0$ and a distance of 
$360 \pm 20$ pc. This agreement with the distance determined from the 
colour-magnitude diagram supports the inference that the companion is 
associated with WASP-31 rather than being a mere line-of-sight neighbour. 
At a distance of 360~pc, the sky-projected separation of the two stars suggests 
that they would be separated by at least 12\,600 AU (0.2 light-year).

Assuming aligned stellar-spin and planetary-orbit axes, the measured \vsini\ of 
WASP-31 and its derived stellar radius indicate a rotational period of 
$P_{\rm rot} = 7.9 \pm 0.7$ d. Combining this with the $B - V$ colour of an 
an F8 star from \citet{2008oasp.book.....G}, we used the relationship of 
\citet{2007ApJ...669.1167B} to estimate a gyrochronological age of 
$950 \pm 250$ Myr. This is an upper limit as the star would be rotating faster, 
and so be younger, than suggested by spectroscopic \vsini\ if the stellar spin 
axis were inclined with respect to the sky plane. 
We used the method of \citet{2011PASP..123..547M} to search for rotational 
modulation of the WASP light curves, as can be caused by the combination of 
magnetic activity and stellar rotation. No evidence of modulation was found. 

The lithium abundance (\ali\ = $2.75 \pm 0.10$) found in WASP-31 implies an 
age \citep{2005A&A...442..615S} between that of open clusters such as M34 
(250 Myr; \ali\ = $2.92 \pm 0.13$) and NGC~752 (2 Gyr; \ali\ = $2.65 \pm 0.13$). 
However, lithium is a poor indicator of age for a star as hot as WASP-31, 
and the measured abundance is consistent at the 1-$\sigma$ level with that of 
the upper envelope of the 5-Gyr M67 (\ali\ = $2.55 \pm 0.18$).

\begin{figure}
\centering                     
\includegraphics[width=84mm]{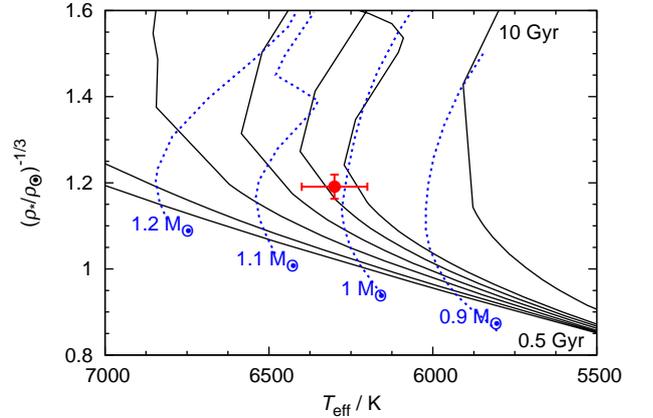}
\caption{Modified H-R diagram. 
The isochrones ($Z = 0.012\approx$ \feh\ = $-$0.20) for the ages 0.5, 1, 2, 3, 
4, 5 and 10 Gyr and the evolutionary mass tracks 
($Z = 0.012 \approx$ \feh\ = $-$0.20; $Y = 0.30$) are from 
\citet{2008A&A...482..883M}.
To obtain the mass tracks, we performed a simple linear interpolation of their  
$Z = 0.0008$ and $Z = 0.017$ tracks.
\label{fig:evol}}
\end{figure}

We interpolated the stellar evolution tracks of \citet{2008A&A...482..883M} 
using \densstar\ from the MCMC analysis and using \teff\ and \feh\ from the 
spectral analysis (Figure~\ref{fig:evol}).  
This suggests an age of $4 \pm 1$ Gyr and a mass of $1.0 \pm 0.1$ \msol\ for 
WASP-31.

Taking account of each age indicator, we suggest a likely age of 
$1^{+3}_{-0.5}$ Gyr. 

\section{Discussion}
With a mass of 0.48 \mjup\ and a radius of 1.55 \rjup, WASP-31b has a density 
13 per cent that of Jupiter and is $\sim$0.3 \rjup\ larger than predicted 
by standard models of irradiated gas giants \citep{2007ApJ...659.1661F}.
Only WASP-17b \citep{2010ApJ...709..159A}, which has a similar mass 
(0.49 \mjup), is known to have a lower density 
\citep[0.06 \densjup,][]{2011MNRAS.416.2108A}. 

With an increasingly large sample of well-characterised planets, we can begin to 
make statistical inferences as to the physical reasons behind their diverse 
natures. \citet{2011MNRAS.410.1631E} showed the radii of 16 of the 18 known 
low-mass (0.1--0.6 \mjup) planets strongly correlate with equilibrium 
temperature and host-star metallicity. 
The calibration of \citet{2011MNRAS.410.1631E} predicts a radius of 1.39 \rjup\ 
for WASP-31b.
In a similar study, but using a different metallicity dependence and treating 
the 74 known Jupiter-mass (0.2--2.5 \mjup) planets, Anderson \& Iro (in prep.) 
also found 
a strong correlation between planetary radius and equilibrium temperature and 
host-star metallicity. 
The calibration of Anderson \& Iro (in prep.) predicts a radius of 1.23 \rjup\ 
for WASP-31b. 
In each case, the predicted radius of WASP-31b is smaller than the measured 
radius ($1.55 \pm 0.05$ \rjup). 

WASP-31 has a similarly low metallicity to WASP-17 
\citep[\feh\ = $-0.19 \pm 0.09$;][]{2010A&A...524A..25T}, thus both WASP-31b and 
WASP-17b could reasonably be expected to have small cores 
\citep{2006A&A...453L..21G, 2007ApJ...661..502B}.
However, this would only somewhat 
explain why the two planets are so large. Both planets are highly irradiated, 
with WASP-17b being more irradiated than  WASP-31b as, despite being in a 
slightly wider orbit ($a = 0.052$ AU), its host star is larger 
(\rstar\ = 1.58 \rsol) and hotter \citep[\teff\ = 6650 K;][]{2011MNRAS.416.2108A}. 
This results in an equilibrium temperature for WASP-17b hotter by 200 K than for 
WASP-31b and, from this, we could expect WASP-17b to be larger than WASP-31b. 
Both planets, though, are larger than predicted by standard models of irradiated 
giant planets \citep[e.g.][]{2007ApJ...659.1661F}, and by the empirical 
relations of \citet{2011MNRAS.410.1631E} and Anderson \& Iro (in prep.). 
Hence, it seems likely that some additional physics, such as Ohmic heating 
\citep{2010ApJ...714L.238B}, is at play. 

The RV data place a stringent upper limit on WASP-31b's orbital eccentricity 
($e < 0.13; 3 \sigma$). 
It is therefore unlikely that tidal heating resulting from the circularisation 
of an eccentric orbit \citep[e.g.][]{2001ApJ...548..466B} was responsible for 
significantly inflating the planet. However, we could happen to be viewing the 
system soon after circularisation occurred and prior to the planet significantly 
contracting. This would have made finding the planet easier due to the greater 
transit depth.

The metallicity of WASP-31 is at the lower end of what may be expected for 
a star of its age in the Solar neighbourhood \citep[at a Galactocentric radius 
of 8.5 kpc;][]{2009A&A...494...95M}. 

\begin{acknowledgements} 
WASP-South is hosted by the South African Astronomical Observatory and we are 
grateful for their ongoing support and assistance. Funding for WASP comes from 
consortium universities and from the UK's Science and Technology Facilities
Council.
M. Gillon acknowledges support from the Belgian Science Policy Office in the 
form of a Return Grant. 
\end{acknowledgements}


\end{document}